\documentclass[lettersize,journal]{IEEEtran}

\usepackage{amsmath,amssymb,amsfonts}
\usepackage{graphicx}
\usepackage{booktabs}
\usepackage{multirow}
\usepackage{tikz}
\usetikzlibrary{arrows.meta,positioning,calc,fit,backgrounds}
\usepackage[hidelinks]{hyperref}
\usepackage{xcolor}

\newcommand{\dB}{\,\mathrm{dB}}
\newcommand{\GHz}{\,\mathrm{GHz}}

\begin{document}

\title{Zero-Shot Frequency Generalization for Radio Map\\
Prediction via Cross-Attention Physics-Residual Learning}

\author{Sajjad Hussain
\thanks{The author is with National University of Sciences and Technology (NUST), Islamabad, Pakistan. Email: sajjad.hussain2@seecs.edu.pk}}

\markboth{Submitted to IEEE Wireless Communications Letters, August 2026}%
{Author \MakeLowercase{\textit{et al.}}: Zero-Shot Frequency Generalization for Radio Map Prediction}

\maketitle

\begin{abstract}
Deep learning models for radio map prediction are trained and tested at the same carriers and cannot serve unseen frequencies. We propose a two-stream network learning a residual over an analytic prior, fusing environment and physics streams by cross-attention. Evaluated at the query frequency, this free-space and knife-edge prior absorbs dominant frequency scaling of pathloss, leaving a residual that varies little across bands. Across 150 ray-traced scenes and four training carriers (1.8--28 GHz), the method reduces RMSE by 35.3\% zero-shot at unseen carriers and scenes, and by 37.6\% at extrapolated 60 GHz, outperforming interpolated 10 GHz.
\end{abstract}


\section{Introduction}
\IEEEPARstart{A}{pathloss} map specifies the large-scale
channel gain from a transmitter to every receiver location in a geographic
area. Accurate radio maps underpin cell-site planning, user association,
device-to-device link scheduling, beam management, and an expanding set of
6G environment-aware functions. Classical statistical models approximate
pathloss as a decaying function of distance and therefore cannot capture
the shadowing, street-canyon guiding, and diffraction that dominate in
dense urban environments. Site-specific ray
tracing resolves these effects accurately at high computational cost motivating learned radio maps. 

RadioUNet established that a UNet can predict outdoor pathloss maps
from a building layout and transmitter location with millisecond
inference~\cite{levie2021radiounet}. Later work added transformer~\cite{tian2021radionet}, generative~\cite{zhang2023rmegan}, graph-based~\cite{chen2023gnn}, and multi-scale feature fusion UNet~\cite{Hussainmsff2026} architectures. More recently, the ICASSP~2023 and~2025 pathloss radio
map prediction challenges have further standardized evaluation of learned
predictors on ray-traced data, spanning outdoor and indoor
propagation. Among indoor
entries, IPP-Net~\cite{feng2025ippnet} showed that injecting a modified 3GPP analytic model as an
auxiliary input to a UNet improves
generalization to unseen geometries.

Most models are trained and tested at a single carrier, and the few multi-frequency studies share same carriers between training and testing. Li \emph{et al.}~\cite{li2024radiogat} reconstructed radio maps jointly across bands by exploiting inter-band correlation; however, their framework requires measurements at the target band. Huai \emph{et al.}~\cite{huai2026crosscity} trained a single model on multiple carriers and validated it across cities, but did not hold out any carrier frequency during training, and therefore did not evaluate cross-frequency generalization. The ICASSP 2025 indoor challenge withheld one carrier, and one of its three tasks explicitly evaluates performance on an unseen carrier. However, the withheld $2.4\GHz$ carrier lies within the $0.868$--$3.5\GHz$ training range, making the task an indoor interpolation problem below $6\GHz$. Zero-shot frequency \emph{extrapolation}, predicting radio maps at carriers above the frequency range observed during training, therefore remains unexplored.

This letter addresses that gap with the following contributions. We
formulate zero-shot frequency generalization and build a ray-traced
benchmark of $150$ urban scenes over six carriers, four for training
and two withheld, one interpolated and one an octave above the training
range\footnote{The codebase and dataset is publicly available at: \url{https://github.com/sajjadhussa1n/PRCA-Net}}. We propose the Physics-Residual with Cross-Attention Network (PRCA-Net), a two-stream encoder--decoder learning a residual over an
analytic prior with the streams fused by cross-attention. We show
this formulation predicts the
extrapolated band more accurately than the interpolated one, a transfer that also holds for a prior-based competitor of different architecture, whereas prior-free learning barely improves on the prior it was meant to surpass.

Rest of the letter is organized as  follows. Sec.~\ref{sec:problem} presents the problem formulation, Sec.~\ref{sec:method} the proposed model,
Sec.~\ref{sec:setup} the dataset, Sec.~\ref{sec:results} the
experimental setup and results, and Sec.~\ref{sec:conclusion}
presents conclusion.

\section{Problem Formulation}\label{sec:problem}
A \emph{deployment} is represented by the triple
$(H,\mathbf{p}_t,f)$: with building-height raster $H\in\mathbb{R}_{+}^{N\times N}$ ($N=256$) at
resolution of $\delta$~m/pixel ($\delta=2$), $\mathbf{p}_t=(x_t,y_t,z_t)$ is the
transmitter position and $f$ is the carrier frequency.
The target
$L_{\mathrm{RT}}\in\mathbb{R}^{N\times N}$ assigns to every receiver
pixel the ray-traced pathloss in dB where a binary mask $M$ excludes pixels
inside buildings and pixels the ray tracer does not reach. 
The task is
to learn a function that predicts the pathloss, $g_{\theta}:(H,\mathbf{p}_t,f)\mapsto \hat{L}$ minimizing masked error against $L_{\mathrm{RT}}$.

Training uses a scene set
$\mathcal{S}_{\mathrm{tr}}$ and a discrete band set
$\mathcal{F}_{\mathrm{tr}}=\{1.8,\,3.5,\,7,\,28\}\GHz$. At test time
the model is queried at $f\notin\mathcal{F}_{\mathrm{tr}}$ with no
gradient update, no fine-tuning, and no calibration measurements at the
new band, i.e. the \emph{zero-shot} setting. 
We further separate the two held-out bands $\{10,\,60\}\GHz$ , because they probe qualitatively
different regimes: $10\GHz$ lies inside the convex hull of
$\mathcal{F}_{\mathrm{tr}}$ (\emph{interpolation}), whereas $60\GHz$ is
more than an octave above the highest training band
(\emph{extrapolation}).

\begin{figure*}[!t]
	\centering
	\resizebox{\textwidth}{!}{%
		\begin{tikzpicture}[
		font=\large,
		>={Stealth[length=2mm]},
		env/.style={rectangle,draw=green!60,fill=green!15,thick,minimum width=1.7cm,minimum height=0.95cm,rounded corners=2mm,align=center},
		phy/.style={rectangle,draw=orange!70,fill=orange!18,thick,minimum width=1.7cm,minimum height=0.95cm,rounded corners=2mm,align=center},
		inp/.style={rectangle,draw=blue!60,fill=blue!15,thick,minimum width=1.9cm,minimum height=1.0cm,rounded corners=3mm,align=center},
		fuse/.style={rectangle,draw=purple!70,fill=purple!18,thick,minimum width=2.0cm,minimum height=1.5cm,rounded corners=2mm,align=center},
		dec/.style={rectangle,draw=red!70,fill=red!15,thick,minimum width=1.7cm,minimum height=0.95cm,rounded corners=2mm,align=center},
		outbox/.style={rectangle,draw=purple!70,fill=purple!12,thick,minimum width=1.9cm,minimum height=1.0cm,rounded corners=3mm,align=center},
		gate/.style={circle,draw=gray!70,fill=gray!12,thick,minimum width=0.6cm,inner sep=1pt,font=\large\bfseries},
		arrow/.style={->,thick},
		down/.style={->,thick,green!50!black},
		skip/.style={->,thick,red!70,dashed},
		block/.style={rectangle,draw=black!45,thick,densely dotted,inner sep=0.30cm,rounded corners=4mm}
		]
		
		\node[inp] (envin) at (0,2.0) {\textbf{Env.\ stream}\\{\small $256^2\!\times\!3$}\\{\small height, Tx, log-dist}};
		\node[inp] (phyin) at (0,-2.0) {\textbf{Phys.\ stream}\\{\small $256^2\!\times\!2$}\\{\small prior, LoS mask}};
		
		\foreach \lvl/\x/\res/\ch in {0/2.6/256/32, 1/5.0/128/64, 2/7.4/64/128, 3/9.8/32/256}{
			\node[env] (e\lvl) at (\x,2.0)  {E\lvl\\{\small $\res^2\!\times\!\ch$}};
			\node[phy] (p\lvl) at (\x,-2.0) {P\lvl\\{\small $\res^2\!\times\!\ch$}};
			\node[gate] (g\lvl) at (\x,0) {$g$};
			\draw[->,gray!60] (e\lvl.south) -- (g\lvl.north);
			\draw[->,gray!60] (p\lvl.north) -- (g\lvl.south);
		}
		\draw[down] (envin) -- (e0);
		\draw[down,orange!60!black] (phyin) -- (p0);
		\foreach \a/\b in {0/1, 1/2, 2/3}{
			\draw[down] (e\a) -- node[font=\small,above,inner sep=1pt]{$\downarrow$} (e\b);
			\draw[down,orange!60!black] (p\a) -- node[font=\small,below,inner sep=1pt]{$\downarrow$} (p\b);
		}
		
		\node[env] (eb) at (12.2,2.0)  {Env BN\\{\small $16^2\!\times\!256$}};
		\node[phy] (pb) at (12.2,-2.0) {Phys BN\\{\small $16^2\!\times\!256$}};
		\draw[down] (e3) -- node[font=\small,above,inner sep=1pt]{$\downarrow$} (eb);
		\draw[down,orange!60!black] (p3) -- node[font=\small,below,inner sep=1pt]{$\downarrow$} (pb);
		
		\node[fuse] (ca) at (14.9,0) {\textbf{Cross-Attn}\\{\small $16^2\!\times\!256$}\\{\small Q:env}\\{\small K,V:phys}};
		\draw[arrow] (eb.east) -- (ca.north west);
		\draw[arrow] (pb.east) -- (ca.south west);
		
		\foreach \lvl/\x/\res/\ch in {3/17.6/32/256, 2/19.8/64/128, 1/22.0/128/64, 0/24.2/256/32}{
			\node[dec] (d\lvl) at (\x,0) {D\lvl\\{\small $\res^2\!\times\!\ch$}};
		}
		\draw[arrow] (ca.east) -- (d3.west);
		\foreach \a/\b in {3/2, 2/1, 1/0}{
			\draw[arrow,red!70] (d\a) -- node[font=\small,above,inner sep=1pt]{$\uparrow$} (d\b);
		}
		
		\foreach \lvl/\drop in {0/-5.2, 1/-4.7, 2/-4.2, 3/-3.7}{
			\draw[skip,rounded corners=3mm]
			(g\lvl.east) -- ++(0.75,0) -- ++(0,\drop) -| (d\lvl.south);
		}
		
		\node[outbox] (resid) at (28.0,0) {\textbf{Residual} $\hat r$\\{\small $256^2\!\times\!1$}};
		\draw[arrow] (d0) -- node[font=\small,above,inner sep=1pt]{Conv$1^2$} (resid);
		\node[inp,fill=orange!12,draw=orange!70] (prioradd) at (31.2,-2.0) {$L_{\mathrm{prior}}$\\{\small (query freq.)}};
		\node[outbox] (final) at (31.2,0) {$\hat L = L_{\mathrm{prior}}+\hat r$};
		\draw[arrow] (resid) -- (final);
		\draw[arrow,orange!70,dashed] (prioradd) -- (final);
		
		\begin{scope}[on background layer]
		\node[block,fit=(envin)(phyin)(eb)(pb),label={[above=0.10cm,font=\large\bfseries]Two-Stream Encoder}] {};
		\node[block,fit=(ca),label={[above=0.10cm,font=\large\bfseries]Fusion}] {};
		\node[block,fit=(d0)(d3)(resid),label={[above=0.10cm,font=\large\bfseries]Decoder}] {};
		\end{scope}
		\end{tikzpicture}}
	\caption{PRCA-Net architecture. Box labels give
		width\,$\times$\,height\,$\times$\,channels. The environment encoder
		(green) uses $[\mathrm{Conv}3^2\!\to\!\mathrm{GN}(8)\!\to\!\mathrm{SiLU}]\times2$
		per stage; the physics encoder (orange) is deliberately lighter, with a
		single $\mathrm{Conv}3^2$. Solid $\downarrow$ denotes a stride-2
		$\mathrm{Conv}4^2$ downsample and $\uparrow$ a bilinear upsample
		followed by $\mathrm{Conv}3^2$. At each resolution a gate
		$g=\sigma(\mathrm{Conv}1^2[\text{env},\text{phys}])$ forms
		$g\cdot\text{env}+(1-g)\cdot\text{phys}$, which is concatenated into the
		decoder along the dashed skip paths. The streams are fused at the
		$16\times16$ bottleneck by cross-attention with environment queries and
		physics keys and values, and the decoded residual $\hat r$ is added to
		the analytic prior evaluated at the query frequency.}
	\label{fig:arch}
\end{figure*}
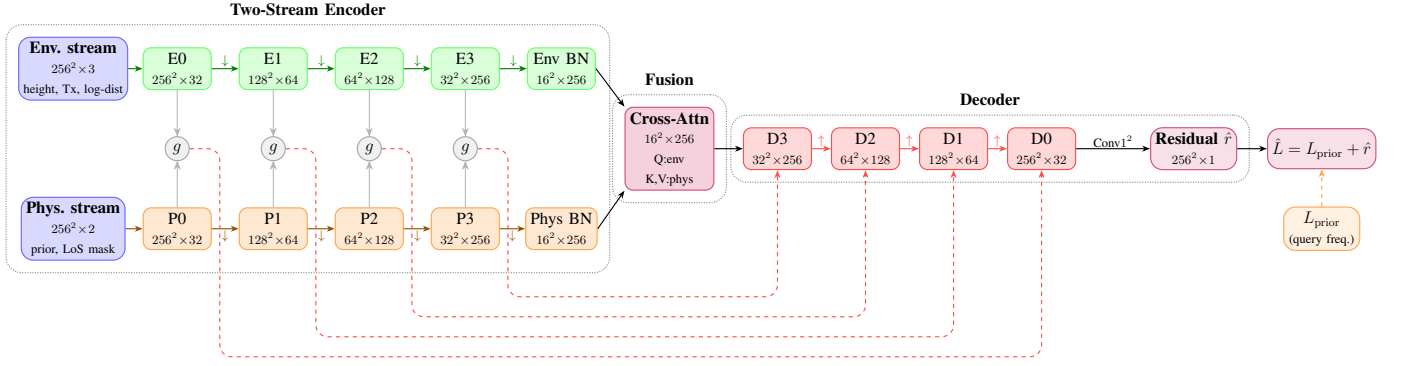
 
\section{Proposed Method}
\label{sec:method}

Our proposed solution consists of (i) a closed-form analytic prior evaluated at
the query frequency, (ii) a residual learning target defined against
that prior, and (iii) a two-stream encoder--decoder that fuses learned
geometry features with the analytic field by \textit{cross-attention}.
Fig.\ref{fig:arch} shows the complete pipeline.

\subsection{Analytic Physics Prior}
\label{sec:prior}
The prior supplies a per-pixel, closed-form pathloss estimate that the
network subsequently corrects. For a transmitter at $\mathbf{p}_t$ and
a receiver pixel at $(x,y,h_r)$, let
$d_3=\lVert \mathbf{p}_t-(x,y,h_r)\rVert_2$ be the three-dimensional
separation. The free-space loss $L_{\mathrm{FSPL}}$ at carrier $f$ is
\begin{equation}
L_{\mathrm{FSPL}} = 20\log_{10}\!\left(\frac{4\pi d_3 f}{c}\right).
\label{eq:fspl}
\end{equation}
Terrain and building obstruction are captured by a simple single knife-edge
diffraction term. Marching along the transmitter--receiver line, we
retain the dominant obstruction, of height excess $h_{ob}$ above the
line-of-sight (LoS) and at along-path distances $d_1$ and $d_2$ from the two
endpoints, we compute the Fresnel--Kirchhoff parameter
\begin{equation}
\nu = h_{ob}\sqrt{\frac{2\,(d_1+d_2)}{\lambda\, d_1 d_2}},
\qquad \lambda = c/f .
\label{eq:nu}
\end{equation}
The diffraction loss $L_{\mathrm{dif}}$ follows the Lee approximation of
ITU-R~P.526~\cite{iturp526}: with $g(\nu)=\sqrt{(\nu-0.1)^2+1}+\nu-0.1$,
\begin{equation}
L_{\mathrm{dif}}(\nu) =
\begin{cases}
0, & \nu \le -0.78,\\[2pt]
6.9 + 20\log_{10} g(\nu), & \nu > -0.78,
\end{cases}
\label{eq:dif}
\end{equation}
applied only where the path is obstructed ($h_{ob}>0$). 
The prior is
$L_{\mathrm{prior}} = \min(L_{\mathrm{FSPL}}+L_{\mathrm{dif}},
L_{\max})$, with the ceiling $L_{max}=160$~dB preventing
unbounded values in deep shadow. 

Crucially, $L_{\mathrm{prior}}$
is evaluated \emph{at the query frequency}, so it carries the dominant
frequency dependence of pathloss analytically. What the prior does \emph{not} capture is multi-bounce reflection,
facade scattering, street-canyon waveguiding, and material-dependent
transmission. The network need only to learn
the frequency-\emph{residual} structure that the prior misses.

\subsection{Inputs and Learning Target}
\label{sec:inputs}
The network consists of two input streams. The \emph{environment stream} is a
three-channel tensor of frequency-invariant geometry consisting of normalised
building height, a transmitter location Gaussian centred on the transmitter pixel, and a log-distance map. The
\emph{physics stream} is a two-channel tensor consisting of the prior
$L_{\mathrm{prior}}$ evaluated at the query frequency, and a
geometry-only LoS mask. 

The learning target is the residual
$r = L_{\mathrm{RT}} - L_{\mathrm{prior}}$, and the prediction is
reconstructed as $\hat{L} = L_{\mathrm{prior}} + \hat{r}$. Training
minimises the masked objective
\begin{equation}
\mathcal{L}(\theta)=\frac{1}{|M|}\sum_{u\in M}
\big|\hat{r}_u - r_u\big|
\;+\;\lambda\,\mathcal{L}_{\mathrm{fc}},
\label{eq:loss}
\end{equation}
an $\ell_1$ loss on the residual, with $M$ the validity mask
(ray-traced, non-building pixels). The second term,
$\mathcal{L}_{\mathrm{fc}}$ ($\lambda=0.1$), is a frequency-consistency
regularizer that penalizes differences between the predicted residual at
different frequencies, evaluated only at pixels that are LoS and valid at
both frequencies of each compared pair. This encourages the residual to be
frequency-flat in LoS regions where propagation is unobstructed, consistent with the
observation that the prior's error is frequency-stable in LoS but grows
with frequency in Non-line-of-sight (NLoS); the residual is left unconstrained in the latter.
Encouraging this regularization helps the network generalize to unseen
carriers.

\subsection{Cross-Attention Physics-Residual Network}
\label{sec:arch}
The backbone of the network is a two-stream encoder--decoder architecture. Both streams are
encoded at four resolutions
($256\!\to\!128\!\to\!64\!\to\!32$, with channel widths $32/64/128/256$). The
environment stream uses two-convolutions per encoder stage while the physics stream uses
single-convolution per encoder stage as
it processes an already-informative analytic signal. At the $16\times16$
bottleneck the two streams are fused by \emph{cross-attention} where the
environment tokens form the queries and physics tokens form the keys and
values. We do not simply concatenate the two
streams. The reliability of $L_{\mathrm{prior}}$ is strongly
space-varying as it is near-exact in LoS regions and degrades where multi-bounce energy dominates. Concatenation
forces a single, spatially fixed weighting of the analytic field.
Cross-attention instead lets every environment token select \emph{where in the analytic field} to draw evidence from, so the network can lean
on the prior where geometry says it is trustworthy and override it where it is not. At each skip resolution the streams are combined by a
lightweight learned gate. A mirrored decoder with gated skip connections
reconstructs the residual $\hat{r}$, which is added to the prior to yield the
predicted map.

\section{Dataset}
\label{sec:setup}

We generate a ray-traced pathloss dataset with Sionna~RT~\cite{sionna}. Fifteen dense urban districts are selected, spanning Manhattan-style grids (NYC Midtown, Chicago Loop, SF Financial, LA Downtown, Barcelona Eixample), European
old-town cores (Hamburg Altstadt, Madrid Sol, Milan Duomo, Rostock, Montpellier Comedie), and mixed mid-rise districts (Berlin Mitte, Munich, Paris, London, Boston Downtown), on average $10$  non-overlapping tiles are extracted per district, for $150$ scene tiles in total. Building footprints and heights are derived from
OpenStreetMap and rasterized to a per-scene height
map on a fixed $256\times256$ grid with each tile covering a square region of
$512$~m per side. 
Pathloss maps are computed with reflection depth
$5$, rooftop diffraction enabled, scattering disabled, and
$3.2 \times 10^{8}$ rays per transmitter, over ITU-R~P.2040 concrete
buildings with frequency-dependent permittivity and
concrete ground, using isotropic antennas at both ends.
For each scene we place $8$ transmitters, sampled across three
deployment strata including \emph{street
	level} ($2$--$10$~m), \emph{rooftop} ($1.5$~m above the local
building), and \emph{mast} ($25$--$35$~m) to reflect realistic deployment. All receivers are
ground-level users at $1.5$~m. This yields $150\times8=1200$
scene--transmitter pairs. Every pair is simulated at four training carriers, $1.8$, $3.5$, $7$,
and $28\GHz$, and at two held-out carriers of $10$ and $60\GHz$. Across all six carriers the dataset comprises
$1200\times6=7200$ ray-traced maps. Scenes are partitioned into training, validation, and test sets by district-wise stratified split of $124/13/13$ scenes
($992/104/104$ pairs). Because all carriers are simulated for every partition, all
six generalization regimes, train/validation/test scene $\times$ train/held-out
frequency, are populated and the scene and frequency axes can
therefore be varied one at a time. The split is frozen to disk and
reused verbatim by every model, including all baselines, to eliminate
partition-induced variance.

\section{Experiments and Results}
\label{sec:results}

\subsection{Training and Evaluation Protocol}
All models are trained for at most $100$ epochs with AdamW (decoupled
weight decay $10^{-4}$) and a base learning rate of $3\times10^{-4}$,
annealed by a closed-form cosine schedule that reaches its floor of
$10^{-6}$ at the final epoch of the budget, with $D_4$ dihedral
augmentation applied to the training split only. Training halts once
validation loss has not improved for $15$ consecutive epochs, and the
checkpoint of lowest validation reconstruction error is retained. Errors are reported as pathloss
RMSE in dB, pooled globally over pixels within each regime rather than
averaged per map, since per-map averaging weights sparsely covered maps
equally with dense ones and biases the statistic. Evaluation spans the
six regimes as discussed above, with the test split --held-out frequency regime
further separated into $10\GHz$ (interpolation) and $60\GHz$
(extrapolation), and LoS and NLoS pixels reported separately for further evaluation.

\subsection{Baselines}
Three baselines are trained and evaluated on the identical frozen
split. 

A per-pixel \emph{Random Forest (RF) Regressor} predicts the residual from the six features including building height, transmitter Gaussian,
log-distance, LoS mask, analytic prior, and normalised frequency with
no spatial context. It isolates how much performance follows
from the input representation alone, and how much from spatial
structure and attention.

\emph{RadioUNet-C (RU-C)}~\cite{levie2021radiounet} is adapted here by supplying the carrier as a normalised
frequency channel alongside the binary building mask and the
transmitter Gaussian as in the original network \cite{levie2021radiounet}. We use cascaded two-stage design where the first
UNet is trained on these three channels alone, then frozen, and its
predicted map is appended as a fourth channel to the input of the
second UNet, which is trained on top of it. 

\emph{IPP-Net (adapted)}~\cite{feng2025ippnet} is a single-stream
UNet that ingests all the same inputs as our model including an analytic prior. The
original network \cite{feng2025ippnet} uses a 3GPP \emph{indoor}-hotspot model, which does not
apply to our outdoor scenes. We substitute the appropriate outdoor
3GPP~TR~38.901 model~\cite{tr38901}, selecting UMi-Street-Canyon or
UMa per transmitter by comparing its height to the local mean rooftop height. This is the closest frequency-aware counterpart to our model.

\subsection{Generalization Across Splits and Bands}
Table~\ref{tab:quadrant} reports pathloss RMSE over the
scene$\times$frequency grid, with the analytic prior included as a
reference. Across the six cells our model stays within 10.28–11.63 dB, against 11.29–12.08 dB for IPP-Net, 13.79–16.54 dB for RF, and 14.94–17.31 dB for RU-C. The two prior-based models occupy a band lying entirely below the prior-free ones. Cross-frequency prediction is thus not a task that established architectures can easily solve. Our model improves on the prior in every cell, by $29.3\%$ to
$41.2\%$, and both generalization axes prove inexpensive. Scene transfer
costs $0.84\dB$: at trained carriers, unseen test scenes reach $11.12\dB$
against $10.28\dB$ on training scenes. Frequency transfer costs less: on test scenes, moving to carriers never seen in training
changes the error by $0.14\dB$, from $11.12$ to $11.26\dB$ and it does
so against a prior that is itself $0.52\dB$ worse at those carriers
($16.87$ to $17.39\dB$), confirming that the held-out bands are the
intrinsically harder ones to model. The baselines separate along the frequency axis according to how much
physics they carry. Across the same shift on test scenes, RF loses $2.29\dB$ and RU-C $0.53\dB$, whereas the two
prior-based models lose $0.20\dB$ (IPP-Net) and $0.14\dB$ (ours). The
IPP-Net transfers nearly as cleanly as our model confirms that zero-shot cross-frequency transfer to be a
property of the physics-residual formulation itself, holding across two
architectures and two different priors, our FSPL and knife-edge
construction, and IPP-Net's 3GPP UMi/UMa model. What the two-stream
cross-attention design adds is a consistent margin on top, $0.38$ to
$1.06\dB$ across the grid, and one that widens in every split when the
carrier is held out. 
The prior-free baselines, by contrast, barely clear
the reference they were meant to surpass: at held-out carriers RF and RU-C come within $0.9\dB$ of the analytic prior
on test scenes ($16.54$ and $16.69\dB$ against $17.39\dB$), and on
validation scenes both are worse than it.

\begin{table}[!t]
	\centering
	\caption{Pathloss RMSE ($\dB$) by scene$\times$frequency regime.}
	\label{tab:quadrant}
	\renewcommand{\arraystretch}{1.2}
	\setlength{\tabcolsep}{3pt}
	\begin{tabular}{@{}llcccccc@{}}
		\toprule
		Scene & Freq. & Prior & RF & RU-C & IPP-Net & \textbf{Ours} & Reduction \\
		\midrule
		Train      & train    & $15.71$ & $13.79$ & $14.94$ & $11.29$ & $\mathbf{10.28}$ & $34.6\%$ \\
		Validation & train    & $16.69$ & $14.56$ & $16.81$ & $11.97$ & $\mathbf{11.59}$ & $30.6\%$ \\
		Test       & train    & $16.87$ & $14.25$ & $16.16$ & $11.77$ & $\mathbf{11.12}$ & $34.1\%$ \\
		Train      & held-out & $17.75$ & $16.27$ & $15.40$ & $11.50$ & $\mathbf{10.44}$ & $41.2\%$ \\
		Validation & held-out & $16.44$ & $16.53$ & $17.31$ & $12.08$ & $\mathbf{11.63}$ & $29.3\%$ \\
		Test       & held-out & $17.39$ & $16.54$ & $16.69$ & $11.97$ & $\mathbf{11.26}$ & $35.3\%$ \\
		\bottomrule
	\end{tabular}
\end{table}

\subsection{Held-Out Carriers and Propagation Condition}
Table~\ref{tab:perfreq} separates the two held-out carriers on unseen
test scenes (last row of Table \ref{tab:quadrant}). At the interpolated $10\GHz$ band our model attains
$11.61\dB$ against the prior's $17.33\dB$, a $33.0\%$ reduction; at
$60\GHz$ extrapolated band, it attains $10.89\dB$
against $17.44\dB$, a $37.6\%$ reduction that confirms that the prior absorbs the dominant frequency-dependent offset analytically, leaving a
residual whose structure is stable across bands. 
IPP-Net likewise
improves under extrapolation ($12.23$ to $11.71\dB$), while our margin
over it widens from $0.62$ to $0.82\dB$.

Splitting by propagation condition shows where the gain originates. The
prior is already accurate in LoS ($4.35$--$4.67\dB$), where free-space
loss plus a single knife-edge term is a reasonable model, and every learned model improves on it only modestly as our
reductions there are $24.1\%$ and $19.3\%$ respectively. The prior loss is poor in NLoS
($18.93$--$19.03\dB$), where one knife-edge term badly under-models
urban multi-diffraction and facade scattering, and it is there that
learning pays. Our model reaches $12.66\dB$ at $10\GHz$ and $11.83\dB$
at $60\GHz$, reductions of $33.1\%$ and $37.8\%$. The reduction over
the prior is therefore largest exactly where the prior is weakest, and
largest of all in the hardest cell of the study, NLoS pixels at an
extrapolated carrier.

The prior-free baselines fail in complementary ways that isolate the
prior's contribution. RF behaves sensibly at $10\GHz$ (LoS $5.54\dB$)
but its LoS error explodes to $28.21\dB$ at $60\GHz$, roughly six times
the prior. 
RU-C
fails differently: its
LoS error is as large as its NLoS error and worsens under extrapolation
($16.57$ to $19.65\dB$). 
The IPP-Net achieves $3.69\dB$ and $4.08\dB$ in LoS and $13.33\dB$ and $12.72\dB$ in NLoS at $10$~GHz and $60$~GHz bands respectively that are close to our model and attributable to the
physics-residual formulation.

\begin{table}[!t]
	\centering
	\caption{Held-out carriers on unseen (test) scenes: RMSE ($\dB)$ overall
		and split by propagation condition, against the analytic prior.}
	\label{tab:perfreq}
	\renewcommand{\arraystretch}{1.2}
	\setlength{\tabcolsep}{3pt}
	\begin{tabular}{@{}llcccccc@{}}
		\toprule
		Band & Cond. & Prior & RF & RU-C & IPP-Net & \textbf{Ours} & Reduction \\
		\midrule
		\multirow{3}{*}{$10\GHz$}
		& All  & $17.33$ & $15.46$ & $16.92$ & $12.23$ & $\mathbf{11.61}$ & $33.0\%$ \\
		& LoS  & $4.35$  & $5.54$  & $16.57$ & $3.69$  & $\mathbf{3.30}$  & $24.1\%$ \\
		& NLoS & $18.93$ & $16.78$ & $16.99$ & $13.33$ & $\mathbf{12.66}$ & $33.1\%$ \\
		\midrule
		\multirow{3}{*}{$60\GHz$}
		& All  & $17.44$ & $17.55$ & $16.45$ & $11.71$ & $\mathbf{10.89}$ & $37.6\%$ \\
		& LoS  & $4.67$  & $28.21$ & $19.65$ & $4.08$  & $\mathbf{3.77}$  & $19.3\%$ \\
		& NLoS & $19.03$ & $14.42$ & $15.72$ & $12.72$ & $\mathbf{11.83}$ & $37.8\%$ \\
		\bottomrule
	\end{tabular}
\end{table}

\section{Conclusion}
\label{sec:conclusion}

We studied zero-shot cross-frequency radio map prediction and showed that
a cross-attention physics-residual network, learning a correction to a
frequency-aware analytic prior, generalizes
to carriers never seen in training, including a $60\GHz$ band well outside
the training range, where it reduces RMSE over the analytic prior by
$37.6\%$. The comparison with no frequency prior baselines and a similar IPP-Net model with a different frequency prior indicate that the inclusion of
frequency information analytically, through a physics prior, is decisive for frequency
generalization. Future work will extend the study to a full
leave-one-frequency-out grid and to additional propagation regimes.

\end{document}